\documentclass[final,2p,times,twocolumn]{elsarticle}
\usepackage[utf8]{inputenc}
\usepackage[english]{babel}
\usepackage{lipsum}
\usepackage{amsmath}
\usepackage{amsfonts}
\usepackage{amssymb}
\usepackage{makeidx}
\usepackage{graphicx}
\usepackage[figuresright]{rotating}
\journal{Materials today communication}

\begin{document}
\begin{frontmatter}
\title{Ab initio study  of structural properties and dynamical stability of C$_3$N$_2$ encapsulating H guest atom }
\author{George S. Manyali } 
\address{$^1$Computational and Theoretical Physics Group, Department of Physical Sciences, Kaimosi Friends University College P.O Box 385-50309, Kaimosi Kenya} 
\fntext[fn1]{gmanyali@kafuco.ac.ke}
\begin{abstract}
We have investigated the structural properties and dynamical stability of C$_3$N$_2$ hosting hydrogen as a guest atom. The calculations were performed using the density functional theory (DFT) as implemented in Quantum espresso code. The generalized gradient approximation approach was used throughout the calculations presented here. We find that the guest atom leaves the lattice constant of the host almost unchanged. We, therefore, noted that C$_3$N$_2$ remains dynamically stable when a single guest atom is inserted into its cage-like structure. We also established that the guest vibrational modes are low-lying modes that lowers the acoustic bandwidth of C$_3$N$_2$.
\end{abstract}
\end{frontmatter}

\section{Introduction}
The cage-like voids in crystal structures plays a key role in modifying   the behavior of materials. The voids are the guest spots where guest atoms are inserted to form a stabilized compound or to improve material's electrical conductivity. A good example are the C$_{60}$ and C$_{20}$ fullerenes. They have a number of very important physical
and chemical properties that make them desirable for special applications.
It is reported that fullerene cages can accommodate more than one guest atoms or molecules \citep{rubin2001insertion,zhang2003thermodynamic}.   Using C$_{20}$ concept, Tian et. al\cite{tian2008superhard} combined carbon and nitrogen to predict C$_3$N$_2$, a novel hard material. This followed other interesting works of Wang et.al \citep{wang2012cagelike}, who reported an unexpected high-pressure stabilization of cagelike diamondoid nitrogen above 263 GPa. This findings provided a significant understanding of behavior of nitrogen related material under extreme conditions. 
Previous studies\citep{wei2015new,HuChenghu2012,du2020theoretical,yuan2021first,kou2021adsorption} on properties of C$_3$N$_2$, did cover the question of whether the cage have inner cavity large enough to hold any atom and whether it remains stable when it hosts a guest atom.

In the present study, atom insertion in the C$_3$N$_2$ host lattice has been performed under zero-pressure conditions. The guest atoms occupied the body-centered vacant site at (0,0,0) in the C$_3$N$_2$ host. We focused on the smallest species with atomic radius(H=0.53\AA). Investigation on guest atoms with atomic radii of 1.12\AA, 1.57\AA, and 2.53\AA did not yield meaningful results and were therefore not considered in this work.

The paper is outlined as follows: In Section 2 we will present the details of our calculations and in Section 3 we will present our results. Finally in Section 4 we will summarize our findings and present our conclusions.

\section{Computational methodology}
The first step in this work was to optimize the geometry of the lattice structures and the determination of
the equations of state. Therefore, the density functional theory (DFT), which has been implemented in the QUANTUM ESPRESSO code\cite{giannozzi2009quantum} was used. Thermo\_pw package \citep{malica2020quasi},  a driver of the Quantum ESPRESSO routines for the calculation of material properties was employed in the calculations of phonons based on density-functional perturbation theory \cite{baroni2001phonons}. The exchange correlation functional known as  Generalized gradient approximation (GGA) in the form of
Perdew-Burke-Ernzerhof (PBE)\citep{perdew1996generalized} and projector augmented wave (PAW) potentials\citep{blochl1994projector} were used. A plane wave cutoff energy of 40Ry, a density cutoff of 300Ry and a 6x6x6 Monkhorst-Pack\citep{monkhorst1976special} grid for the electronic integrations were used throughout. Phonon frequencies were calculated on a 3x3x3 q-point grid.


\section{Results and discussion}
\subsection{Structural properties}
\begin{figure}[h!]
(a)\includegraphics[width=3.5cm]{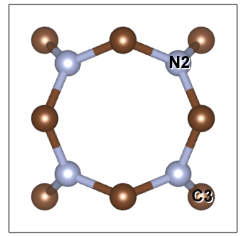}
(b)\includegraphics[width=3.5cm]{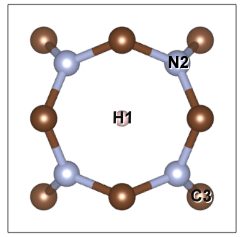} 
\caption{(a)Unit cell of the simple cubic C$_3$N$_2$ with space group Pm-3m(221). The guest atoms occupied the body-centered vacant site at (0,0,0) in the C$_3$N$_2$ host as shown in (b).}
\label{fig:1}       
\end{figure}

The initial configuration of atoms in the unit cell of the simple cubic C$_3$N$_2$ is shown in figure \ref{fig:1}. The guest atom occupied the body-centered vacant site at (0,0,0) in the C$_3$N$_2$ host before the structural optimization of the lattice. After the optimization, the lattice parameter of C$_3$N$_2$ was determined to be 5.095\AA, which was in good agreement with previous studies\cite{tian2008superhard,HuChenghu2012}. When H atom was inserted in the C$_3$N$_2$ host lattice, the cage expands slightly
 as presented in Table\ref{tab:strus}. Such very small distortions may have very little impact on the overall structural and elastic properties. It should be noted that the cage in C$_3$N$_2$ is not large enough to hold atoms such as beryllium, lithium and barium, which have atomic radii of 1.12\AA, 1.57\AA, and 2.53\AA ~ respectively. The guest atom forces the cage to expand, just as the structure loses its dynamical stability as the lattice parameter increases significantly. 

Other parameters such as density also remained largely unchanged, indicating none of the atoms were incorporated into C$_3$N$_2$ host. 
\begin{table}[h!]
\caption{Lattice constant a(\AA), ground-state energy (E0 in Ry) and density $\rho$ (in g/cm$^3$)}
\label{tab:strus}
\begin{tabular}{lll}
Property&C$_3$N$_2$&HC$_3$N$_2$ \\\hline
a&5.095,5.085$^a$,5.049$^b$   &5.103  \\
E0&-446.448096638 &-447.351685595      \\
$\rho$   &3.2159&3.2129  \\\hline
\end{tabular}\\
$^a$Ref.\cite{HuChenghu2012}
$^b$Ref.\cite{tian2008superhard}
\end{table}


\subsection{Phonon dispersion}
\begin{figure}[h!]
\includegraphics[width=4.0cm]{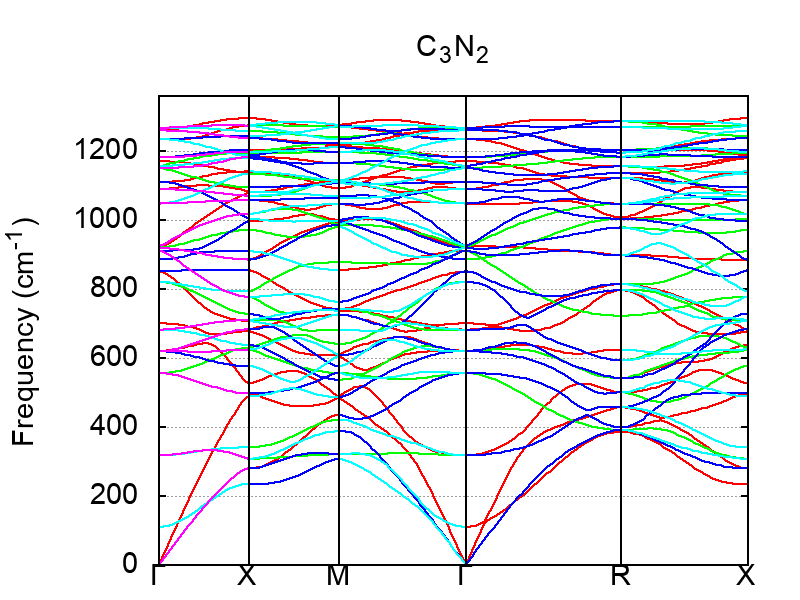}
  \hskip -1.1ex
\includegraphics[width=4.0cm]{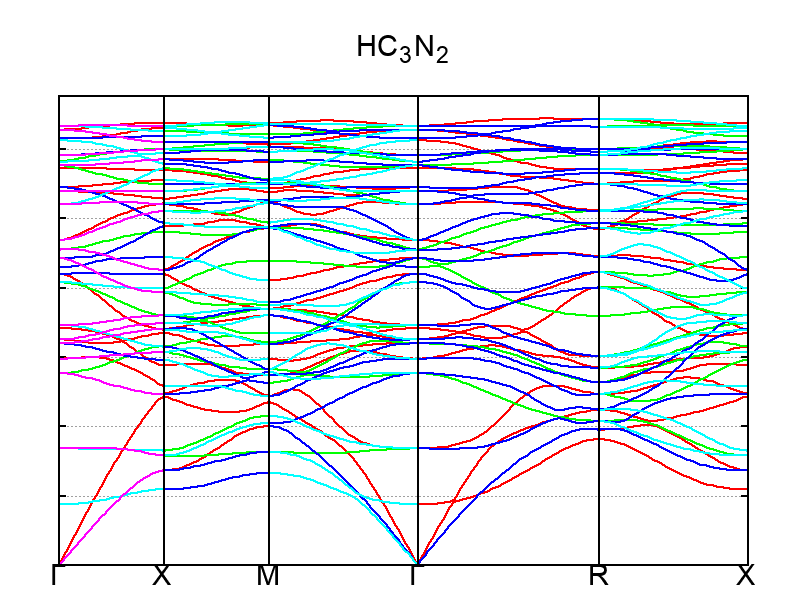} 
\caption{The phonon-dispersion curves for C$_3$N$_2$ (left panel) and  HC$_3$N$_2$ (right panel).}
\label{fig:phon}       
\end{figure}

Our calculated dispersion curves are shown in the figure \ref{fig:phon}. The absence of imaginary modes in the dispersion curves, clearly demonstrates that insertion of H atom in C$_3$N$_2$ does not interfere with dynamical stability. If we Look at the two regions, i.e  one below 400cm$^{-1}$ that encompasses the acoustic modes and the region of optical modes above 1200cm$^{-1}$, we observe very important information.

We find that H guest atom in this C$_3$N$_2$ host, have low-lying vibrational modes, so that their contributions to the heat capacity, the entropy, and the free energy are negligible. It is observed that the guest vibrational modes lowers the bandwidth of the acoustic range. We also note that above 1200cm$^{-1}$, the optic branches are flat, an indication that these modes do not contribute to heat transport in C$_3$N$_2$.

%
%
%

\subsection{Conclusion}
In summary, the structural and dynamical properties of caged-C$_3$N$_2$ encapsulating hydrogen atoms were studied by first-principles calculations. The voids in C$_3$N$_2$ are large enough to accommodate the H atoms, hence the host remains dynamically stable after insertion of the guest atom. The phonon frequencies of the guest atom are within the acoustic range, and are  expected to have little impact on the heat capacity and entropy.

\section*{Acknowledgment}
This work was financially supported by the Kenya Education Network (KENET) through the Computational Modeling and Materials Science (CMMS) Research mini-grants 2019. We acknowledge the Centre for High Performance Computing (CHPC), Cape Town, South Africa, for providing us with computing facilities.

\end{document}